\documentclass{article}
\usepackage[utf8]{inputenc}
\usepackage{amsfonts}
\usepackage{amsmath}
\usepackage{amssymb}
\usepackage{graphicx}
\usepackage{tikz}
\usepackage{multirow}
\usepackage{indentfirst}
\usepackage{titling}
\usepackage{mathrsfs}
\usepackage{graphicx}
\usepackage{cite}
\usepackage{authblk}
\usepackage{titlesec}
\newtheorem{thm}{Theorem}
\newtheorem{ex}{Example}
\providecommand{\keywords}[1]
{
  \textbf{\textit{Keywords---}} #1
}

\title{Latent Variable Model for Multivariate Data with Measure-specific Sample Weights and Its Application in Hospital Compare}
\author[1]{Chengan Du}
\author[2]{Shu-Xia Li}
\author[1,2]{Zhenqiu Lin}
\author[3*]{Haiqun Lin}
\affil[1]{School of Medicine, Yale University , New Haven, Connecticut.}
\affil[2]{Yale New Haven Health Services Corporation/Center for Outcomes Research and Evaluation, New Haven, Connecticut.}
\affil[3*]{School of Nursing, Rutgers University, Newark, New Jersey.}
\affil[ ]{\textit {\{haiqun.lin\}@rutgers.edu}}
\date{July, 2019}                     
\begin{document}
\maketitle
\begin{abstract}

We developed a single factor model with measure-specific sample weights for
multivariate data with multiple observed indicators clustered within a higher level
subject. The factor is therefore a latent variable shared by multiple indicators within a same subject and the sample weights are different across different indicators and different subjects. Even after integrating out the latent variable, the likelihood of the data cannot be written as the sum of weighted likelihood of each subject because a subject has different sample weights respectively for its multiple indicators. In addition, the number of
available indicators varies across subjects. We derive a pseudo likelihood for the latent
variable model with measure-specific weights. We investigate various statistical
properties of the latent variable model with measure-specific sample weights and its
connection to the traditional factor analysis. We found that the latent variable
model provides consistent estimates for its variances when the measure-specific sample
weights are properly re-scaled. Two estimation procedures are developed - EM algorithm
for the pseudo likelihood and marginalization of the pseudo likelihood by directly
integrating out the latent variable to obtain the parameter estimates. This approach is
illustrated by the analysis of publicly reported hospitals with indicators and sample
weights. Numerical studies are conducted to investigate the influence of weights and
their sample distribution.
\end{abstract}


\section{Introduction}
This work is built on the basis of factor analysis. The factor analysis is a widely used statistical tool in many fields, such as psychology, educational testing, social behavior and biomedical sciences \cite{henderson1975best,thompson2007factor}. The factor analysis is popular since it provides a
convenient modeling tool for multiple observed indicators within a subject \cite{muthen2013new}. In this paper, we propose a latent variable model with a
single factor for multivariate data with measure-specific weights that vary across
indicators and across subjects. A pseudo likelihood approach is developed for our
model. \\

Over the years, Centers for Medicare and Medicaid Services (CMS) Hospital Compare website publishes hospitals performance scores which are called hospital indicators in this paper.
CMS hopes that these indicators will help people choose their hospitals. In 2016, CMS started to report the star ratings of more than four thousand hospitals across the whole country \cite{venkateshoverall}. The goal of the CMS overall hospital quality star rating is to estimate one summary score using a total of fifty-seven hospitals indicators collected from the hospital compare database. The fifty-seven indicators are divided into seven different groups according the quality aspect they represent. Each indicator within a hospital has its sample weight representing the volume of patients that contribute to that indicator. A
group-specific factor score is derived for the indicators within the group. The goal of
this paper is to estimate the factor model within a group incorporating sample weights
for each indicator within each hospital.\\

The factor is an unobserved latent variable that represents the underlying
hospital performance. In addition to the presence of measure-specific weights that
vary at both indicator and hospital level, there is a missing data issue as only a few
hospitals report the complete set of all the hospitals indicators. Traditional factor
analysis using correlation matrix approach is not possible to deal with such situation
and therefore we propose a pseudo likelihood method to estimate such model.\\



Existing literature has dealt with subject-specific weights. \cite{shwartz2008estimating} studied the volume related weights which are subject-specific via the hierarchical logistic models.
\cite{veiga2014use} applied subject-specific weights in multivariate multilevel models to the longitudinal data. \cite{landrum2000analytic} gave an approach that based on the  likelihood to generalize the overall score. And \cite{Agostinelli2013} proposed a weighted latent likelihood method based on subject-specific weights. To the best of our knowledge, there are no studies for the latent variable model with measure-specific weights, as well as its asymptotic behaviors. Therefore, we 
fill the gap by proposing a version of weighted pseudo likelihood that fits for the measure-specific weights through two algorithms: Expectation-Maximization (EM) method and the
marginalization of the pseudo likelihood to get the parameter estimates. We apply this model to the CMS hospital compare dataset. \\


The sum of weights for each indicator across the hospitals is set to be the sample size of that indicator so that a hospital with a smaller volume for that indicator has a smaller sample weight for that indicator comparing to a hospital with a larger volume for that indicator. However, we show in Section 3 that the sum of such intuitive sample weights across hospitals for the indicator need to be bounded below the sample size in order for the estimates of the variance for the latent variable to be
consistent. We impose such bound by multiplying each sample weight by 0.99 which in
practice remains the same interpretation.\\

The rest of the paper is organized as follows. In Section 2 we present our model and specify the pseudo likelihood. The statistical properties of the latent variable
model are given in Section 3. In Section 4 we describe the two algorithms including
the EM approach and the marginal likelihood approach. The bound of the weights
is given in Section 3. In Section 5 we conduct the numerical studies. In Section
6 we analyze three datasets from US hospital compare and in Section 7 we conclude
with a discussion.

\section{Model Specification and Pseudo Likelihood}

We start the model with the following set-up:

Suppose there are a total of $m$ indicators with $H$ hospitals (subjects) in each indicator to be evaluated. Let $Y_{jh}$ denote the $j$th indicator in the hospital $h$ with $j=1, \dots, m$ and $h=1, \dots, H$. Let $w_{jh}$ denote the measure-specific weight of hospital $h$ and indicator $j$. For each $h$, we fit a single confirmatory factor model as:
\begin{align} \nonumber
& \quad \quad \quad \quad \quad Y_{jh}|\alpha_h \sim N(\mu_j+\gamma_j\alpha_h, \sigma_j^2) \\ \nonumber
&\text{ with measure-specific weight } w_{jh}, \quad j=1, \dots, m,
\end{align}

where $\alpha_h \sim N(0,1)$ is the underlying factor or latent variable representing hospital $h$'s performance based on its all $m$ indicators. The higher the value of
$Y_{jh}$, the better the performance of hospital h in indicator m. The $\mu_j, \gamma_j$ and $\sigma_j^2$ are unknown parameters that we need to estimate. 


\subsection{The Pseudo Joint Likelihood of Data and Latent Variable}
Given $\alpha_h$, $Y_{1h}, \dots, Y_{mh}$ are conditionally independent. We have the joint density for the latent variable and $Y_{1h},\dots, Y_{mh}$ satisfies  
\begin{align}\nonumber
P(Y_{1h},\dots, Y_{mh},\alpha_h)&=P(\alpha_h)P(Y_{1h},\dots, Y_{mh}|\alpha_h)
\\ \nonumber&=P(\alpha_h)\prod_{j=1}^mP(Y_{jh}|\alpha_h).
\end{align}
We define the joint pseudo likelihood for hospital $h$ with sample weights as
\begin{align}\label{yi}
P^*(Y_{1h},\dots, Y_{mh},\alpha_h)&=P(\alpha_h)\prod_{j=1}^m[P(Y_{jh}|\alpha_h)]^{w_{jh}},
\end{align}
where $w_{jh}$ bounded differentiable non-negative (the sample weight) function \cite{Agostinelli2013} independent to $Y_{jh}$.\\

The logarithm of the term within the product is given as :
\begin{align}\nonumber
\log P^*(Y_{jh}|\alpha_h)&=w_{jh}\log P(Y_{jh}|\alpha_h) \\ \nonumber
&=-w_{jh} \log \sigma_j -\frac{w_{jh}}{2\sigma_j^2}(Y_{jh}-\mu_j-\gamma_j\alpha_h)^2-\frac{w_{jh}}{2}\log 2\pi.
\end{align}
Thus, the conditional log-density of expression \eqref{yi} is given as:
\begin{align}\nonumber
& \quad \quad \log P^*(Y_{1h},\dots, Y_{mh}|\alpha_h)=\sum \limits_{j=1}^mw_{jh}\log P(Y_{jh}|\alpha_h)\\ 
\label{czhang}&=-\sum \limits_{j=1}^m w_{jh} \log \sigma_j -\sum \limits_{j=1}^m \frac{w_{jh}}{2\sigma_j^2}(Y_{jh}-\mu_j-\gamma_j\alpha_h)^2-\frac{1}{2}\sum \limits_{j=1}^m w_{jh} \log 2\pi \\ \nonumber 
&= - \big[\sum \limits_{j=1}^m \frac{w_{jh}\gamma_j^2}{2\sigma_j^2}\alpha_h^2-\sum \limits_{j=1}^m\frac{2w_{jh}(Y_{jh}-\mu_j)\gamma_j}{2\sigma_j^2}\alpha_h+ \sum \limits_{j=1}^m \frac{w_{jh}(Y_{jh}-\mu_j)^2}{2\sigma_j^2}\big]\\ \nonumber
&\quad -\sum \limits_{j=1}^m w_{jh} \log \sigma_j -\frac{1}{2}\sum \limits_{j=1}^m w_{jh} \log 2\pi.
\end{align}

Therefore, the negative logarithm of the joint density of all $m$ indicators for hospital $h$ is
\begin{align}\nonumber
 &-\log P^*(Y_{1h},\dots, Y_{mh},\alpha_h)=-\log P^*(Y_{1h},\dots, Y_{mh}|\alpha_h)-\log P(\alpha_h)\\ \nonumber
&=(\frac{1}{2}+\sum \limits_{j=1}^m \frac{w_{jh}\gamma_j^2}{2\sigma_j^2})\alpha_h^2-\sum \limits_{j=1}^m\frac{2w_{jh}(Y_{jh}-\mu_j)\gamma_j}{2\sigma_j^2}\alpha_h+\sum \limits_{j=1}^m w_{jh} \log \sigma_j \\ \nonumber
&\quad +\sum \limits_{j=1}^m \frac{w_{jh}(Y_{jh}-\mu_j)^2}{2\sigma_j^2}+\frac{1}{2}(\sum \limits_{j=1}^m w_{jh}+1) \log 2\pi\\ \nonumber
&=(\frac{1}{2}+\sum \limits_{j=1}^m \frac{w_{jh}\gamma_j^2}{2\sigma_j^2}) \left \{\alpha_h-\frac{\sum \limits_{j=1}^m\frac{w_{jh}(Y_{jh}-\mu_j)\gamma_j}{\sigma_j^2}}{1+\sum \limits_{j=1}^m \frac{w_{jh}\gamma_j^2}{\sigma_j^2}}\right \}^2+\frac{1}{2}(\sum \limits_{j=1}^m w_{jh}+1) \log 2\pi\\ 
& \label{cdu} \quad \quad \quad +\sum \limits_{j=1}^m w_{jh} \log \sigma_j+\sum \limits_{j=1}^m \frac{w_{jh}(Y_{jh}-\mu_j)^2}{2\sigma_j^2}-\frac{(\sum \limits_{j=1}^m\frac{w_{jh}(Y_{jh}-\mu_j)\gamma_j}{\sigma_j^2})^2}{2+2\sum \limits_{j=1}^m \frac{w_{jh}\gamma_j^2}{\sigma_j^2}}.
\end{align}

And the log joint density of all $m$ indicators for all $H$ hospitals is the summation of $\log P^*(Y_{1h},\dots, Y_{mh},\alpha_h)$ through $1$ to $H$.\\

Note that we can also bring missing values of $Y_{jh}s$ into \eqref{cdu} by setting the corresponding $w_{jh}=0$, therefore, the joint pseudo log-density of latent variable model is also compatible with missing data in $Y$.

\subsection{The Marginal Pseudo Likelihood}
Note that part of \eqref{cdu} can be rewritten as the log-density of a normal distribution:
\begin{align}\nonumber
\eqref{cdu}&=(\frac{1}{2}+\sum \limits_{j=1}^m \frac{w_{jh}\gamma_j^2}{2\sigma_j^2}) \left \{\alpha_h-\frac{\sum \limits_{j=1}^m\frac{w_{jh}(Y_{jh}-\mu_j)\gamma_j}{\sigma_j^2}}{1+\sum \limits_{j=1}^m \frac{w_{jh}\gamma_j^2}{\sigma_j^2}}\right \}^2+\frac{1}{2}\log 2\pi+\frac{1}{2} \log ({1}+\sum \limits_{j=1}^m \frac{w_{jh}\gamma_j^2}{\sigma_j^2})^{-1}\\ \nonumber
& -\frac{1}{2} \log ({1}+\sum \limits_{j=1}^m \frac{w_{jh}\gamma_j^2}{\sigma_j^2})^{-1}+\sum \limits_{j=1}^m w_{jh} \log \sigma_j+\sum \limits_{j=1}^m \frac{w_{jh}(Y_{jh}-\mu_j)^2}{2\sigma_j^2}-\frac{(\sum \limits_{j=1}^m\frac{w_{jh}(Y_{jh}-\mu_j)\gamma_j}{\sigma_j^2})^2}{2+2\sum \limits_{j=1}^m \frac{w_{jh}\gamma_j^2}{\sigma_j^2}}\\ \label{liao}
&+\frac{1}{2}\sum \limits_{j=1}^m w_{jh} \log 2\pi.
\end{align}

The first line of \eqref{liao} is exactly the density function for a normal distribution after logarithm. Denote $Y_{.h}=[Y_{1h},\dots,Y_{mh}]'$ as the indicator vector for hospital $h$, by integration with respect to $\alpha_h$, we have the marginal pseudo log-likelihood (denoted by $\mathcal{L}^*$) of all the parameters for hospital $h$ satisfies 
\begin{align}\nonumber
 &-2\log \mathcal{L}^*(\mu_1 \dots \mu_m, \gamma_1 \dots \gamma_m, \sigma_1 \dots \sigma_m|Y_{.h}) =\sum \limits_{j=1}^m w_{jh} \log 2\pi \\ \label{lin}
 & +\log ({1}+\sum \limits_{j=1}^m \frac{w_{jh}\gamma_j^2}{\sigma_j^2})+\sum \limits_{j=1}^m w_{jh} \log \sigma_j^2+\sum \limits_{j=1}^m \frac{w_{jh}(Y_{jh}-\mu_j)^2}{\sigma_j^2}-\frac{(\sum \limits_{j=1}^m\frac{w_{jh}(Y_{jh}-\mu_j)\gamma_j}{\sigma_j^2})^2}{1+\sum \limits_{j=1}^m \frac{w_{jh}\gamma_j^2}{\sigma_j^2}}.
\end{align} 

\section{ Statistical Properties of the Model}
\subsection{Main Theorems}
In this subsection, we describe the asymptotic behaviors of the latent variable model under uniform weight case (all $w_{jh}=1$) and the varying weights case.\\

We start with the simple uniform weight case when all the weights equal to one. Without missing value, the negative marginal logarithm likelihood (denoted by $\mathcal{L}$) for all the parameters for hospital $h$ satisfies 
\begin{align}\nonumber
 &-2\log \mathcal{L}(\mu_1 \dots \mu_m, \gamma_1 \dots \gamma_m, \sigma_1 \dots \sigma_m|Y_{.h})=m\log 2\pi \\ \label{qqq}
 & +\log ({1}+\sum \limits_{j=1}^m \frac{\gamma_j^2}{\sigma_j^2})+\sum \limits_{j=1}^m  \log
 \sigma_j^2+\sum \limits_{j=1}^m \frac{(Y_{jh}-\mu_j)^2}{\sigma_j^2}-\frac{(\sum \limits_{j=1}^m\frac{(Y_{jh}-\mu_j)\gamma_j}{\sigma_j^2})^2}{1+\sum \limits_{j=1}^m \frac{\gamma_j^2}{\sigma_j^2}},
\end{align} 
since  all $w_{jh}=1$. The pseudo log-likelihood becomes log-likelihood.\\

The asymptotic behaviors of the latent variable model are followed by the result of a toy example:

\begin{ex}
 Let $m=3$, $Z_{jh} \sim N(\mu_j, \sigma_j^2+\gamma_j^2),  j=1,\dots,3, h=1,\dots,H.$ And the covariance between $Z_{i.}=[Z_{i1},\dots,Z_{iH}]$ and $Z_{j.}=[Z_{j1},\dots,Z_{jH}]$ is $\gamma_i\gamma_j, (1 \le i, j \le 3).$ Without missing values, we have twice the negative log-likelihood of $Z$ is the same as \eqref{qqq} when $m=3$.
\end{ex}

Example 1 shows in the latent variable model with uniform weight, When there are no missing values, the latent variable model is the same as the confirmatory factor analysis with single factor, since they both have the same log-likelihood. However, compared to obtaining parameter estimates through the
calculation of the inverse variance-covariance matrix in the confirmatory factor
model with multivariate normal distribution, the estimation through the
likelihood of latent variable model formulation for the factor model is easier to obtain with $m>3$ cases. Therefore, we can have the following results.

\begin{thm}
 When there are no missing values, as $H \to \infty$, we have the expected negative marginal logarithm likelihood ($\text{ENMLL}_H$) satisfies
\begin{align}\nonumber
\text{ENMLL}_H & \xrightarrow{a.s.} \frac{1}{2} \log \big[ (1+\sum \limits _{j=1}^m \frac{\gamma_j^2}{\sigma_j^2})\prod_{j=1}^m \sigma_j^2 \big]+\frac{m}{2}(\log 2\pi+1),
\end{align}
and $\sqrt{H}(\frac{1}{H}\sum \limits_{h=1}^H \text{NMLL}_h-\text{ENMLL}_H)$ has a normal distribution with mean equals to zero, finite variance. 
\end{thm}

Theorem 1 gives the asymptotic behavior of the marginal likelihood of latent variable model with uniform weights.

\begin{thm}
Let the weight matrix $W=[w_{jh}]_{m\times H}$ where $w_{jh}=w_j$, are positive constants. Then by definition of NMLL, we have the expected pseudo marginal log-likelihood (ENWMLL) for hospital $h$ satisfies
\begin{align}\nonumber
    2\text{ENWMLL}_H &\xrightarrow{a.s.} m+\log \big[(1+\sum \limits _{j=1}^m\frac{w_j\gamma_j^2}{\sigma_j^2})\prod_{j=1}^m\frac{\sigma_j^2}{w_j} \big]+\sum\limits_{j=1}^m  (w_j-1)\log  \sigma_j^2\\ \nonumber
    &+\sum \limits_{j=1}^m  \log w_j+\sum \limits_{j=1}^m (w_{j}-1) \log 2\pi,
\end{align}
without missing, as $H \to \infty$, and the central limit theorem also holds for NWMLL.
\end{thm}

Theorem 2 gives the asymptotic behavior of marginal likelihood of the latent variable model with the weights those are specifically assigned.

\subsection{Variance Bounded From Zero}
This subsection mainly address the issue that in the latent variable model, certain $\sigma_j^2$ may become zero. In that case, the validity of the latent variable
model can be endangered. However, the $\sigma^2_j$ can be bounded away from zero by
adjusting the weights in a simple way presented in this section.
\\   


Easy to observe that each $\sigma_j$ should be bounded away from zero in order to make the negative marginal logarithm (pseudo) likelihood valid. In the mean time, the computational speed for estimating the parameters will slow down heavily at the area where certain $\sigma^2$ is tiny since there are no closed form solutions for \eqref{lin}. In the following part, we discuss the approaches that prevent the estimated standard error from going to zero when we incorporate with varying weights.

\subsubsection{Uniform Weight}
We will focus on three indicators ($m=3$) since if the number of indicators exceeds three, we can still pick three indicators to study.

For $j=1,2,3$, let $Y_{jh} \sim N(\mu_j, \sigma_j^2+\gamma_j^2)$, where $h=1,\dots, H$, and the covariance between $Y_{i.}$ and $Y_{j.}$ is $\gamma_i\gamma_j>0, (1 \le i, j \le 3)$. Assume $Y_{1.}, Y_{2.}, Y_{3.}$ have the same variance, moreover, assume $$\text{Corr}(2,3)=\gamma_2\gamma_3/\sqrt{(\gamma_2^2+\sigma_2^2)(\gamma_3^2+\sigma_3^2)}$$ is the smallest correlation. Then we have $\sigma_1^2$ be the smallest parameter among all $\sigma$s since $\gamma_1$ is the largest one among all three $\gamma$s. And, when all the weights equal to one in latent variable model, two extreme examples may cause $\sigma^2_1$ to be exactly zero:

\begin{ex} Assume $Y_{1.}$ and $Y_{2.}$ are identical, then we have $\sigma_1=\sigma_2=0$.
\end{ex}

\textbf{Proof:} Easy to verify that 
$$\text{Corr}(1,2)=\frac{\gamma_1\gamma_2}{\sqrt{\sigma_1^2+\gamma_1^2}\sqrt{\sigma_2^2+\gamma_2^2}}=\frac{\gamma_1^2}{\gamma_1^2+\sigma_1^2}=1.$$

Then we have $\sigma_1=\sigma_2=0$.\\

\begin{ex} Assume $Y_{1.}, Y_{2.}$ and $Y_{3.}$ satisfies Corr(1,2)$\times$Corr(1,3) $>$ Corr(2,3), then we have $\sigma_1^2 = 0$.
\end{ex}

\textbf{Proof:} By the result in Example 1, we have

$$\text{Corr}(1,2)\times \text{Corr}(1,3)=\frac{\gamma_1^2}{\sigma_1^2+\gamma_1^2}\text{Corr}(2,3)\le \text{Corr}(2,3),$$

thus the latent variable model outputs $\sigma_1=0$ as its minimized occupation.\\

When Example 2 or Example 3 happens, the likelihood estimates of the latent variable model will be at the boundary, thus the posterior variance of $\alpha_h$ may become zero. In order to prevent it, we need proper weights to get the posterior variance bounded from zero.\\

\subsubsection{Varying Weights}\label{ccc}
Followed by Theorem 2, let  $W=[w_{jh}]_{m\times H}$ where $w_{jh}=w_j$, are positive constants. We have, 
\begin{align}\nonumber
    2\text{ENWMLL}_H &\xrightarrow{a.s.} m+\log \big[(1+\sum \limits _{j=1}^m\frac{w_j\gamma_j^2}{\sigma_j^2})\prod_{j=1}^m\frac{\sigma_j^2}{w_j} \big]+\sum\limits_{j=1}^m  (w_j-1)\log  \sigma_j^2\\ \nonumber
    &+\sum \limits_{j=1}^m  \log w_j+\sum \limits_{j=1}^m w_{j} \log 2\pi,
\end{align}

as $H \to \infty$. We can see that if we set the sum of weight equals to the sample size ($\bar{w}_{j.}=w_j=1$), we have $\text{ENWMLL}_H = \text{ENLL}_H$
since all the weights are equal to one. Both Example 2 and Example 3 can cause $\sigma$ being zero.\\

Assuming there are not identical indicators among $Y_{1.}, \dots , Y_{m.}$, there is at most one $\sigma$ can be zero. Without loss of generality, assume $\sigma_1^2$ is the smallest among all $\sigma^2$s, then we have 
\begin{align}\nonumber
    |\log \big[(1+\sum \limits _{j=1}^m\frac{w_j\gamma_j^2}{\sigma_j^2})\prod_{j=1}^m\frac{\sigma_j^2}{w_j} \big]| &= |\log\big[( \gamma_1^2 \prod_{j=2}^m\frac{\sigma_j^2}{w_j})+\frac{\sigma_1^2}{w_1}(1+\sum \limits _{j=2}^m\frac{w_j\gamma_j^2}{\sigma_j^2})\prod_{j=2}^m\frac{\sigma_j^2}{w_j} \big]| \\ \nonumber
    &< |\log( \gamma_1^2 \prod_{j=2}^m\frac{\sigma_j^2}{w_j})|< \infty.
\end{align}

Thus we have for $M=\sum \limits_{j=1}^m  \log w_j+\sum \limits_{j=1}^m w_j \log 2\pi$ which is a constant,
\begin{align}\nonumber
    2\text{ENWMLL}_H-M& \xrightarrow{a.s.} m+\sum\limits_{j=2}^m  (w_j-1)\log  \sigma_j^2+(w_1-1)\log\sigma_1^2\\ \label{sss}
    &+\log\big[( \gamma_1^2 \prod_{j=2}^m\frac{\sigma_j^2}{w_j})+\frac{\sigma_1^2}{w_1}(1+\sum \limits _{j=2}^m\frac{w_j\gamma_j^2}{\sigma_j^2})\prod_{j=2}^m\frac{\sigma_j^2}{w_j} \big] \\ \nonumber
    &\xrightarrow{a.s.} m+\sum\limits_{j=2}^m  (w_j-1)\log  \sigma_j^2+\log( \gamma_1^2 \prod_{j=2}^m\frac{\sigma_j^2}{w_j})+(w_1-1)\log\sigma_1^2.
\end{align}

If we have $w_1-1<0$, it will penalize the expected marginal weighted likelihood from $\sigma_1$ being zero since both $(w_1-1)\log\sigma_1^2 \to \infty$ will hold, and the rest terms are bounded.\\

Furthermore, if we let 
\begin{align}\nonumber
   S_3= ( \gamma_1^2 \prod_{j=2}^m\frac{\sigma_j^2}{w_j})/\big[\frac{1}{w_1}(1+\sum \limits _{j=2}^m\frac{w_j\gamma_j^2}{\sigma_j^2})\prod_{j=2}^m\frac{\sigma_j^2}{w_j}\big]>0,
\end{align}

by the Dominated Convergence Theorem, taking derivative to \eqref{sss} with respect to $\sigma_1^2$ yields
\begin{align}\nonumber
   \frac{\partial{\text{ENWMLL}_H}}{\partial{\sigma_1^2}}\xrightarrow{a.s.}\frac{1}{2(\sigma_1^2+S_3)}+\frac{w_1-1}{2\sigma_1^2}<0
\end{align}
at the neighborhood larger than zero for $\sigma_1^2$ when $w_1-1<0$. Which proves that $\sigma_1^2$ is bounded from zero.\\

Similarly, if we have $w_1>1$, then $\text{ENWMLL}_H \to -\infty$ as $\sigma_1$ approaching zero. And $\sigma_1=0$ will become an optimal estimate since the pseudo likelihood then goes to infinity. Therefore, we showed that if we have $w_j<1, j=1, \dots, m$, then we can ensure all the estimated standard errors are bounded from zero.\\

Both numerical study and data analysis will show that by setting the mean of weights smaller than one, under the case which $W=[w_{jh}]_{m\times H}$ denotes the weight matrix with arbitrary values, we still can have $\sigma$s bounded from zero property.\\ 

\section{Estimation}
Two approaches (EM \cite{dempster1977maximum} and the marginal) will be provided in this section.

\subsection{The EM Algorithm}

\subsubsection{E-Step:} 

Since \eqref{cdu} has an exact form of normal distribution, we can get the posterior mean of $\alpha_h$ to be
$$x_h=E(\alpha_h|Y_{.h},\mu,\gamma,\sigma^2)=\frac{\sum \limits_{j=1}^m\frac{w_{jh}(Y_{jh}-\mu_j)\gamma_j}{\sigma_j^2}}{1+\sum \limits_{j=1}^m \frac{w_{jh}\gamma_j^2}{\sigma_j^2}},$$ 
and its posterior variance is
$$y_h=Var(\alpha_h|Y_{.h},\mu,\gamma,\sigma^2)=({1}+\sum \limits_{j=1}^m \frac{w_{jh}\gamma_j^2}{\sigma_j^2})^{-1},$$

by the definition of normal pdf. Along with these, we also need to calculate the posterior second moment in our EM approach: 
\begin{align}\nonumber
z_h=E(\alpha_h^2|Y_{.h},\mu,\gamma,\sigma^2)&=E^2(\alpha_h|Y_{.h},\mu,\gamma,\sigma^2)+Var(\alpha_h|Y_{.h},\mu,\gamma,\sigma^2)\\ \nonumber
&=({1}+\sum \limits_{j=1}^m \frac{w_{jh}\gamma_j^2}{\sigma_j^2})^{-1}+\left \{\frac{\sum \limits_{j=1}^m\frac{w_{jh}(Y_{jh}-\mu_j)\gamma_j}{\sigma_j^2}}{1+\sum \limits_{j=1}^m \frac{w_{jh}\gamma_j^2}{\sigma_j^2}}\right \}^2
\end{align}
given $Y_{jh}, \mu_j, \gamma_j$ and $\sigma_j^2, \quad j = 1,\dots, m, h=1,\dots, H$ .\\

\subsubsection{M-Step:}

Note that directly minimizing \eqref{cdu} has computational difficulty, an alternative way is to maximize \eqref{czhang} under the condition where $\alpha_h=x_h, h=1,\dots,H$, and repeat the process in E-step.\\

We adopt an iterative method by firstly taking derivatives to \eqref{czhang} of all $H$ hospitals with respect to $\mu_j, \gamma_j$ and $\sigma_j,$ $j = 1,\dots, m$. \\

Therefore, our iterative method in M step is 

\begin{align*}
\hat{\mu}_j&= \mathop{\mathrm{argmax}}\limits_{\mu_j} \sum \limits_{h=1}^Hw_{jh}\log P(Y_{jh}|x_h,z_h) \\ 
&=\sum \limits_{h=1}^H w_{jh}(Y_{jh}-\gamma_jx_h)/\sum \limits_{h=1}^H {w_{jh}}, \\
\hat{\gamma}_j&= \mathop{\mathrm{argmax}}\limits_{\gamma_j} \sum \limits_{h=1}^Hw_{jh}\log P(Y_{jh}|x_h,z_h) \\
&=\sum \limits_{h=1}^H {w_{jh}}x_h(Y_{jh}-\mu_j)/\sum \limits_{h=1}^H {w_{jh}z_h}
\end{align*}

\begin{align}\nonumber
&\text{and} \quad \hat{\sigma}_j^2 = \mathop{\mathrm{argmax}}\limits_{\sigma_j} \sum \limits_{h=1}^Hw_{jh}\log P(Y_{jh}|x_h,z_h) =\\ \nonumber
&\sum \limits_{h=1}^H w_{jh}\{(Y_{jh}-\mu_j)^2-2(Y_{jh}-\mu_j)\gamma_jx_h+\gamma_j^2z_h\}/\sum \limits_{h=1}^H w_{jh}, \quad j=1,\dots,m.
\end{align}

Once the Expectation-Maximization algorithm converges, we can update the latent variables $\alpha_h$ by
$\frac{\sum \limits_{j=1}^m\frac{w_{jh}(Y_{jh}-\hat{\mu}_j)\hat{\gamma}_j}{\hat{\sigma}_j^2}}{1+\sum \limits_{j=1}^m \frac{w_{jh}\hat{\gamma}_j^2}{\hat{\sigma}_j^2}}, h=1,\dots,H$. We repeat this procedure several times until every $\alpha_h$ becomes stable.\\

In the M-step, the solutions of $\hat{\mu}$s, $\hat{\gamma}$s and $\hat{\sigma}^2$s are consistent regardless the choices for any initial values since 
\begin{align}\nonumber
    \frac{\partial^2 \eqref{cdu} }{\partial \mu_j^2} &=\sum \limits _{h=1}^H\frac{w_{jh}}{\sigma_j^2}>0, \\ \nonumber
    \frac{\partial^2 \eqref{cdu} }{\partial \gamma_j^2} &=\sum \limits _{h=1}^H\frac{w_{jh}\alpha_h^2}{\sigma_j^2}>0, \quad j=1,\dots, m.
\end{align}

In the same time, the values of $x_h$ in the E-step also maximize the joint pseudo log-likelihood with respect to $\alpha$, and
\begin{align}\nonumber
\frac{\partial^2 \eqref{cdu} }{\partial \alpha_h^2} &=1+\sum \limits _{h=1}^H\frac{w_{jh}\alpha_h^2}{\sigma_j^2}>0, \quad h=1,\dots,H,
\end{align}

suggests that \eqref{cdu} is convex for all the $\mu$s, $\gamma$s and $\alpha$s. \\

Moreover, if we assume $\sigma_j^2 \ge \epsilon>0$ holds for some positive number of $\epsilon$, then  
\begin{align}\label{condition}
   {\sigma}_j^2<2\sum \limits _{h=1}^Hw_{jh}(Y_{jh}-\mu_j-\gamma_j\alpha_h)^2 /\sum \limits _{h=1}^Hw_{jh}
\end{align}

holds for every indicator. Since by the result in M-step, the estimated value of $\sigma_j^2$ is closed to $\sum \limits _{h=1}^Hw_{jh}(Y_{jh}-\mu_j-\gamma_j\alpha_h)^2 /\sum \limits _{h=1}^Hw_{jh}.$  \eqref{condition} implies

\begin{align}\nonumber
    \frac{1}{{\sigma}_j^2}\sum \limits _{h=1}^Hw_{jh}(Y_{jh}-\mu_j-\gamma_j\alpha_h)^2 > \frac{1}{2}\sum \limits _{h=1}^Hw_{jh}
\end{align}
holds. Thus we have at $(0,\epsilon)$,
\begin{align}\nonumber
\frac{\partial^2 \eqref{cdu} }{\partial (\sigma_j^2)^2} &=-\sum \limits _{h=1}^H \frac{w_{jh}}{\sigma_j^4}+2\sum \limits _{h=1}^H\frac{w_{jh}}{\sigma_j^6}(Y_{jh}-\mu_j-\gamma_j\alpha_h)^2 \\ \nonumber
&=\frac{1}{\sigma_j^4}\sum \limits _{h=1}^H w_{jh}\big[\frac{2(Y_{jh}-\mu_j-\gamma_j\alpha_h)^2}{\sigma_j^2}-1\big]>0.
\end{align}

 Therefore, the EM approach is the same as the coordinate descent method, we can find the minimizer, for $\sigma_j^2 \in (0,\epsilon)$. \cite{daubechies2004iterative}\\

In addition, if there are more than two indicators that contribute to the latent variable, then we have the local maximum should be the global maximum for \eqref{cdu}. \cite{yong2013beginner}

\subsection{The Marginal Pseudo Likelihood}
Given $W$ as constant, conditional on $Y$, we can get the parameter estimates directly by maximizing \eqref{lin}, i.e.
\begin{align}\nonumber
 &-2\log \mathcal{L}^*(\mu_1 \dots \mu_m, \gamma_1 \dots \gamma_m, \sigma_1 \dots \sigma_m|Y_{.h}) =\sum \limits_{j=1}^m w_{jh} \log 2\pi \\ \nonumber
 & +\log ({1}+\sum \limits_{j=1}^m \frac{w_{jh}\gamma_j^2}{\sigma_j^2})+\sum \limits_{j=1}^m w_{jh} \log \sigma_j^2+\sum \limits_{j=1}^m \frac{w_{jh}(Y_{jh}-\mu_j)^2}{\sigma_j^2}-\frac{(\sum \limits_{j=1}^m\frac{w_{jh}(Y_{jh}-\mu_j)\gamma_j}{\sigma_j^2})^2}{1+\sum \limits_{j=1}^m \frac{w_{jh}\gamma_j^2}{\sigma_j^2}},
\end{align} 
then use the estimates to get all $\alpha_h$s by
$$\hat{\alpha}_h=E(\alpha_h|Y_{.h},\mu,\gamma,\sigma^2)=\frac{\sum \limits_{j=1}^m\frac{w_{jh}(Y_{jh}-\mu_j)\gamma_j}{\sigma_j^2}}{1+\sum \limits_{j=1}^m \frac{w_{jh}\gamma_j^2}{\sigma_j^2}},$$ 
and their posterior variances by
$$\hat{Var}(\alpha_h)=Var(\alpha_h|Y_{.h},\mu,\gamma,\sigma^2)=({1}+\sum \limits_{j=1}^m \frac{w_{jh}\gamma_j^2}{\sigma_j^2})^{-1},$$
where $1 \le h \le H$. We implement this algorithm through the NLMIXED procedure in SAS. 

\section{Numerical Study}

In this section, we study two cases based on measure-specific weights. The weights in every indicator are set to be mean equal to one for both cases at beginning. For every $j$, the weights satisfy: $w_{jh}$ are exponentially shaped (the extreme case) and $w_{jh}$ are mounded shaped (the regular case) for $h=1,\dots,H.$ We show that under the regular case, the latent variable gets consistent results where those results have been incorporated with the weight information. Under the extreme case, we demonstrate that by setting the sum of the weights less than the sample size in each indicator, or, equivalently, mean of each weights to be less than one, the estimates of $\sigma$s can be always bounded from zero. We can then get both the variables estimates as well as the variances estimates for all the latent variables.
This is consistent with the results in section \eqref{ccc}. We also compare the performance of the latent variable model with different sample sizes through our algorithm in the extreme case.

\subsection{A Regular Case}\label{bbb}
Assume there are three indicators in the group, let the sample size $H=1000$, the inputs $Y_{i.}=[Y_{i1},\dots, Y_{iH}]$ $,i=1,2,3$ are generated through a multivariate normal distribution with mean zero and variance one, the correlation among each pair of two indicators is set as $0.5$.\\

We firstly generate $H$ random numbers with Gamma distribution $G(3/2, 1/2)$ to get $W_{1.}=[w_{11},\dots, w_{1H}]$, where $3/2$ is the shape parameter, and $1/2$ is the scale parameter. For $W_{2.}$ and $W_{3.}$, we separately generate $H$ random numbers with Gamma distribution of $G(3, 1/3)$. Then we divide the $W_{i.}, i=1,2,3$ by their sample means, respectively. Thus we get the sample mean of the weights of every indicator be the same, one.\\

We replicate the study based on the above setting for $100$ times, and focus on both average loading ($\gamma$) and average standard deviation ($\sigma$) for the three indicators in the result. For contrast, we also replicate the study for uniform weights $100$ times. Table 1 shows the results:

\begin{table}
\caption{Parameter Estimates With Varies Weights VS Uniform Weights}
\centering
\begin{tabular}{lllllllll}
	\hline \hline
	& & \multicolumn{2}{c}{Weight} & & & \multicolumn{2}{c}{No Weight}&\\
\hline
 $\gamma_1$ & 0.8520 & $\sigma_1$ & 0.4801 & $\gamma_1$ & 0.7084 & $\sigma_1$ & 0.7057
\\
 $\gamma_2$ & 0.6675 & $\sigma_2$ & 0.7391 & $\gamma_2$ & 0.7080 & $\sigma_2$ & 0.7088
\\ 
 $\gamma_3$ & 0.6700 & $\sigma_3$ & 0.7356 & $\gamma_3$ & 0.7107 & $\sigma_2$ & 0.7053
\\ 
\hline \hline
\end{tabular}
\end{table}

We can see with uniform weights, by design, all three indicators seem to have similar average loadings and average root mean squared errors. By Theorem 1, the result with uniform weights is the same as the confirmatory factor analysis with one factor. When there are weights which are moderately skewed, the average loading for the first indicator is larger than those of the second and the third. This is meaningful since although all three weights have the same mean, $W_{1.}$ is more skewed than $W_{2.}$ and $W_{3.}$. The maximum value in $W_{1.}$ is larger than those of $W_{2.}$ and $W_{3.}$. Thus the latent variable model can also output the result with different weights.

\subsection{An Extreme Case}
We use the same generating method for $Y_{jh}$ as section \eqref{bbb}, where $j=1,2,3, h=1,\dots,H.$
And we use Gamma distributions with the shape parameters are smaller than the scale parameters to generate more skewed weights.\\

 We firstly generate $H$ random numbers with Gamma distribution of $G(1/2, 2)$ to get $W_{1.}.$ For $W_{2.}$ and $W_{3.}$, we separately generate $H$ random numbers with Gamma distributions of $G(1, 2)$. Again, we divide the $W_{i.}, i=1,2,3$ by their sample means, respectively. We will find $\sigma_1 \to 0$ as section 5.2 suggests, thus we focus on the result of $\sigma_1$ here. To make comparison, we test the performances of latent variable model under $0.9*W, 0.8*W, 0.7*W$ and $0.99*W$.\\
 
\begin{figure}
  \centering
      \includegraphics[scale=0.6]{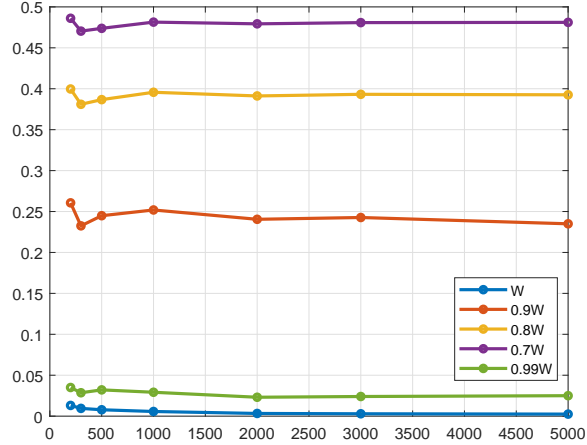}
  \caption{Values of ARMSE1 among different setting of weights for increasing sample size}
\end{figure}
 
We compare the performance of all five weight matrices with sample size $H$ varies from $300$ to $5000$, we replicate the study $100$ times under each value of the sample size. Figure 1 shows the average values of $\sigma_1$.\\

Figure 1 shows that, with the original weights of means equal to one, the smallest average root mean squared error (ARMSE1) tends to be zero. However, the smaller the means of weights of the indicators, the larger the ARMSE1. As the sample size goes larger, the result of ARMSE1 tends to be stable. Therefore, if we have the mean of weights of the indicator to be smaller than one, we will prevent the $\sigma$s from being zero, then we can get the estimates of the posterior variances of the latent variables.

\section{Data Analysis}

We applied our latent variable model to the CMS’s Overall Hospital Quality Rating database from the CMS 2019 public data across the subjected States. This database consists seven indicator groups: Mortality; Readmission; Safety of Care (Safety); Patient Experience; Effectiveness; Timeliness and Image Efficiency. In this section, we first analyze two indicator groups in the three outcome groups: Mortality and Readmission. We will then discuss the group of Safety.\\

In each indicator group, hospital had the reported indicator
scores. For each indicator, the scores from the available
hospitals are standardized with mean zero and variance one. There also exist measure-specific weights (CMS calls
them as the denominator weights) for the hospitals reflecting their volumes of
admissions. Similar as sample weights, the mean of the denominator weights in
every indicator is standardized as just below one. Note that those weights vary
across both indicator level and hospital level, therefore, the latent variable model
is appropriated for the data.

\subsection{Mortality (regular)}

For the group of mortality, seven indicators among $4573$ hospitals are presented:

1. MORT-30-AMI: Acute Myocardial Infarction (AMI) 30-Day Mortality Rate;

2. MORT-30-CABG: Coronary Artery Bypass Graft (CABG) 30-Day Mortality Rate;

3. MORT-30-COPD: Chronic Obstructive Pulmonary Disease (COPD) 30-Day Mortality Rate;

4. MORT-30-HF: Heart Failure (HF) 30-Day Mortality Rate;

5. MORT-30-PN: Pneumonia (PN) 30-Day Mortality Rate;

6. MORT-30-STK: Acute Ischemic Stroke (STK) 30-Day Mortality Rate;

7. PSI-4-SURG-COMP: Death Among Surgical Patients with Serious Treatable Complications.\\

We apply both the EM approach and the marginal approach to the mortality data, and calculate the maximum absolute value of predicted the latent variables, the difference is only 2.2208e-04. This suggests that the EM and the marginal approach are identical. Table 2 shows the parameter estimates of the latent variable model. We found that the loadings are balanced across indicators, all the estimated variances are bounded from zero. This result is the same as the result from CMS via the SAS quadrature method. \cite{venkateshoverall} 

\begin{table}
\caption{Parameter Estimates in the Mortality Group with Un-adjusted Weights}
\centering
\begin{tabular}{llllll}
	\hline \hline
	$\mu$ & Un-adj  & $\gamma$ & Un-adj  & $\sigma$ & Un-adj \\
\hline
$\mu_1$ & 0.113 & $\gamma_1$ & 0.508 & $\sigma_1$ & 0.927
\\
$\mu_2$ & 0.131 & $\gamma_2$ & 0.333 & $\sigma_2$ & 0.894
\\ 
$\mu_3$	& 0.002 & $\gamma_3$ & 0.676 & $\sigma_3$ & 0.822
\\ 
$\mu_4$	& 0.107 & $\gamma_4$ & 0.713 & $\sigma_4$ & 0.682
\\  
$\mu_5$ & -0.007 & $\gamma_5$ & 0.665 & $\sigma_5$ & 0.740
\\
$\mu_6$ & -0.049 & $\gamma_6$ & 0.484 & $\sigma_6$ & 0.975
\\ 
$\mu_7$	& -0.061 & $\gamma_7$ & 0.281 & $\sigma_7$ & 1.049
\\ 
\hline \hline
\end{tabular}
\end{table}

We also multiplied $0.99$ to all the weights in the mortality data, and we found there is no difference in the parameter estimates between $0.99*W$ and $W$. Moreover, the rooted mean square error of the latent variable $\alpha_h$ between the original weight and $0.99$ multiples the weight method is $0.0025$. Therefore, the $0.99$ times weights performs very closely to the method with un-adjusted weights in the mortality group.

\subsection{Readmission (Extreme)}
In the data of the readmission group, there are nine indicators among $4573$ hospitals:

1. EDAC-30-AMI: Excess Days in Acute Care (EDAC) after hospitalization for Acute Myocardial Infarction (AMI);

2. EDAC-30-HF: Excess Days in Acute Care (EDAC) after hospitalization for Heart Failure (HF);

3. EDAC-30-PN: Excess Days in Acute Care (EDAC) after hospitalization for Pneumonia (PN);

4. OP-32: Facility 7-Day Risk Standardized Hospital Visit Rate after Outpatient Colonoscopy;

5. READM-30-CABG: Coronary Artery Bypass Graft (CABG) 30-Day Readmission Rate;

6. READM-30-COPD: Chronic Obstructive Pulmonary Disease (COPD) 30-Day Readmission Rate;

7. READM-30-Hip-Knee: Hospital-Level 30-Day All-Cause Risk-Standardized Readmission Rate (RSRR) Following Elective Total Hip Arthroplasty (THA)/Total Knee Arthroplasty (TKA);

8. READM-30-HOSP-WIDE: HWR Hospital-Wide All-Cause Unplanned Readmission;

9. READM-30-STK: Stroke (STK) 30-Day Readmission Rate.\\

 After running the latent variable model, we found the estimated $\sigma_8$ for the 30 day hospital-wide readmission indicator is zero. This is because in the indicator of 30 day hospital-wide readmission, the numbers of admissions varies from $25$ to $23915$, which are way larger than the rest indicators in the group. After standardization, the distribution of denominator weights are skewed much more heavily in 30 day hospital-wide readmission than the rest indicators. Thus we apply the method of $0.99$ times the weights to the 30 day hospital-wide readmission indicator, in order to force its standard error larger than zero, as well as keep the parameter estimates as close as possible. The result of parameter estimates is shown in Table 3. 

\begin{table}
\caption{Parameter Estimates in the Readmission Group via Un-adjusted and Adjusted Weights}
\centering
\begin{tabular}{lllllllll}
	\hline \hline
	$\mu$ & Un-adj & Adj  & $\gamma$ & Un-adj & Adj & $\sigma$ & Un-adj & Adj\\
\hline
$\mu_1$ & 0.031 & 0.031 & $\gamma_1$ & 0.316 & 0.318 & $\sigma_1$ & 0.710 & 0.710
\\
$\mu_2$ & -0.175 & -0.175 & $\gamma_2$ & 0.427 & 0.430 & $\sigma_2$ & 0.748 & 0.747
\\ 
$\mu_3$	& -0.243 & -0.243 & $\gamma_3$ & 0.410 & 0.413 & $\sigma_3$ & 0.775 & 0.774
\\ 
$\mu_4$	& 0.198 & 0.198 & $\gamma_4$ & -0.002 & -0.002 & $\sigma_4$ & 1.228 & 1.228
\\  
$\mu_5$ & 0.106 & 0.106 & $\gamma_5$ & 0.303 & 0.304 & $\sigma_5$ & 1.024 & 1.024
\\
$\mu_6$ & -0.068 & -0.067 & $\gamma_6$ & 0.522 & 0.525 & $\sigma_6$ & 0.972 & 0.971
\\ 
$\mu_7$	& 0.194 & 0.194 & $\gamma_7$ & 0.388 & 0.390 & $\sigma_7$ & 1.043 & 1.042
\\ 
$\mu_8$	& 0.000 & 0.001 & $\gamma_8$ & 0.975 & 0.978 & $\sigma_8$ & 0.000 & 0.056
\\ 
$\mu_9$	& -0.051 & -0.051 & $\gamma_9$ & 0.499 & 0.502 & $\sigma_9$ & 0.983 & 0.982
\\ 
\hline \hline
\end{tabular}
\end{table}

We can see except for $\sigma_8$, all the parameters from the original weight and the adjusted weight methods have difference less than 0.001. Moreover, the 0.99 adjusted method can ensure all the variance in the readmission group bounded from zero. This is consistent with previous numerical and theoretical results. 

\subsection{Safety (Extreme)}
Modeling the group of Safety has been challenging over the years by its unbalanced loadings and bi-peak parameter estimates through the latent variable modeling.\\

In the Safety of Care group, there are eight indicators among $4573$ hospitals:

1. COMP-HIP-KNEE: Hospital-Level Risk-Standardized Complication Rate (RSCR) Following Elective Primary Total Hip Arthroplasty (THA) and Total Knee Arthroplasty (TKA);

2. HAI-1: Central-Line Associated Bloodstream Infection (CLABSI);

3. HAI-2: Catheter-Associated Urinary Tract Infection (CAUTI);

4. HAI-3: Surgical Site Infection from colon surgery (SSI-colon);

5. HAI-4: Surgical Site Infection from abdominal hysterectomy (SSI-abdominal hysterectomy);

6. HAI-5: MRSA Bacteremia;

7. HAI-6: Clostridium Difficile (C. difficile);

8. PSI-90-Safety: Complication/Patient Safety for Selected Indicators (PSI).\\

Similar to the readmission group, the safety group is also an extreme case since both COMP-HIP-KNEE and PSI-90-Safety indicators have much larger variance in numbers of admissions than the rest six indicators. After running the latent variable model with the un-adjusted sample weights, we found the loadings for HAI-1 to HAI-6 are closed to zero. This will cause the bi-peak issue of the parameter estimates since
there are only two indicators loaded unto the latent factor, one or both of $\sigma_1$ or $\sigma_8$ can have zero variance. We provided two sets of methods to solve the issue in the safety group.  

\subsubsection{Solution One}
One way to solve the problem is comparing the marginal pseudo log-likelihood between the two peaks. And we found when PSI-90-Safety ($\sigma_8$) turned out to be zero, the marginal log-likelihood is larger. Similarly as the readmission group, we apply the methods of $0.99$ times weights (denoted by Adj) to both COMP-HIP-KNEE and PSI-90-Safety indicators. Table 4 shows the parameter estimates between the original weights and $0.99$ times weights in the Safety group. And Table 5 shows the root mean squared errors of the latent variable estimates between $0.9$, $0.8$, $0.7$ multiply weights method and the method of multiplying $0.99$ to the un-adjusted weights (since under the method of un-adjusted weights, the latent variable variances cannot be calculated). \\

\begin{table}
\caption{Parameter Estimates in the Safety of Care Group via Different Weightings}
\centering
\begin{tabular}{lllllllll}
	\hline \hline
	$\mu$ & Un-adj & Adj  & $\gamma$ & Un-adj & Adj & $\sigma$ & Un-adj & Adj\\
\hline
$\mu_1$ & 0.287 & 0.287 & $\gamma_1$ & 0.188 & 0.189 & $\sigma_1$ & 1.039 & 1.039
\\
$\mu_2$ & -0.007 & -0.007 & $\gamma_2$ & 0.007 & 0.007 & $\sigma_2$ & 0.723 & 0.723
\\ 
$\mu_3$	& -0.010 & -0.010 & $\gamma_3$ & 0.008 & 0.008 & $\sigma_3$ & 0.757 & 0.757
\\ 
$\mu_4$	& -0.055 & -0.055 & $\gamma_4$ & 0.045 & 0.046 & $\sigma_4$ & 0.837 & 0.837
\\  
$\mu_5$ & 0.010 & 0.010 & $\gamma_5$ & 0.060 & 0.060 & $\sigma_5$ & 0.867 & 0.867
\\
$\mu_6$ & 0.032 & 0.032 & $\gamma_6$ & 0.037 & 0.037 & $\sigma_6$ & 0.796 & 0.796
\\ 
$\mu_7$	& 0.003 & 0.003 & $\gamma_7$ & 0.025 & 0.025 & $\sigma_7$ & 0.622 & 0.622
\\ 
$\mu_8$	& 0.016 & 0.016 & $\gamma_8$ & 0.897 & 0.901 & $\sigma_8$ & 0.000 & 0.033
\\ 
\hline \hline
\end{tabular}
\end{table}

\begin{table}
\caption{Summary Statistics Under Different Weights}
\centering
\begin{tabular}{lllll}
	\hline \hline
	 & 0.99*W & 0.9*W & 0.8*W & 0.7*W \\
	 \\
\hline
 RMSE & & 0.0861 & 0.2123 & 0.3552 \\
 Mean & 0.0000 & 0.0000 & 0.0000 & 0.0000 \\
 Std  & 0.8354 & 0.7711 & 0.6754 & 0.5604 \\
\hline \hline
\end{tabular}
\end{table}

We found that the parameters estimate are very closed between the original weight and adjusted weight, and as the weight coefficient decreases from 0.99 to 0.7, the standard error of the predicted latent variables is getting farther away from the prior variance (one). 

\subsubsection{Solution Two}
Another idea of modeling the Safety group is smoothing the weights thus prevent the dominance of either COMP-HIP-KNEE  and  PSI-90-Safety. Consider the HAIs are similar in both admission volumes and indicator scores, this method is seeking a balanced loading from the latent variable model.

\begin{table}
\caption{Parameter Estimates in the Safety Group via log transformation}
\centering
\begin{tabular}{llllll}
	\hline \hline
	$\mu$ & log-W  & $\gamma$ & log-W  & $\sigma$ & log-W \\
\hline
$\mu_1$ & 0.048 & $\gamma_1$ & 0.105 & $\sigma_1$ & 0.995
\\
$\mu_2$ & 0.036 & $\gamma_2$ & 0.528 & $\sigma_2$ & 0.738
\\ 
$\mu_3$	& 0.016 & $\gamma_3$ & 0.372 & $\sigma_3$ & 0.846
\\ 
$\mu_4$	& 0.000 & $\gamma_4$ & 0.211 & $\sigma_4$ & 0.919
\\  
$\mu_5$ & 0.022 & $\gamma_5$ & 0.279 & $\sigma_5$ & 0.897
\\
$\mu_6$ & 0.033 & $\gamma_6$ & 0.359 & $\sigma_6$ & 0.867
\\ 
$\mu_7$	& 0.019 & $\gamma_7$ & 0.078 & $\sigma_7$ & 0.865
\\ 
$\mu_8$	& 0.008 & $\gamma_8$ & 0.134 & $\sigma_8$ & 0.938
\\ 
\hline \hline
\end{tabular}
\end{table}

One option is taking logarithm transformation to the admission volume for all indicators in the Safety group. This will help to reduce the variance in skewness of the un-adjusted weights in the Safety group. The result is in Table 6. We can see that the loadings are balanced and there are more than three indicators with relatively high loadings in the latent variable. Thus taking logarithm to the admission volume in the Safety group can help make the result balanced and thus identifiable.

\section{Summary and Discussion}
We present a latent variable model that incorporates measure-specific sample weights via pseudo-likelihood estimation in this work. The latent variable model can handle the missing value issue as well.
The estimates obtained through the algorithms have desirable asymptotic properties. We gave examples in both numerical study and real data analysis where the latent variable model can produce zero standard error estimates for certain indicators. We showed that if the sample weights means of those indicators are less than one, the estimates of variance components are bounded away from zero. We provided a log transformation method prior to the sample weights can help to obtain nonzero variance components as well.\\

For future work, we plan to investigate the pseudo likelihood and the estimating algorithm of the latent variable model under random weights with Gamma distribution, thus we can discover the threshold between the choice of the shape (scale) parameters and the bounded-away-from-zero estimates of variance components. Moreover, we would also like to investigate more behaviors of the latent variable model under varies distributions of weights, such as Poisson distribution, beta distribution, etc.


\section{Appendix}
We organize this section as: \eqref{7.1} and \eqref{7.2} are the proof to Example 1, \eqref{7.3} is the proof to Theorem 1, \eqref{7.4} is the proof to Theorem 2.\\

Recall that For any hospital $h$, there are $m$ indicators. Assume there exists an overall latent score $\alpha_h$ from the $m$ indicators, conditional on its score,  each indicator has an independent normal distribution with
$$Y_{jh}|\alpha_h \sim N(\mu_j+\gamma_j\alpha_h, \sigma_j^2), \quad \quad 1\le j\le m$$
where $Y_{jh}$ is the response for the $j$th indicator in hospital $h$. $\mu_j, \gamma_j$ and $\sigma_j^2$ are unknown parameters. \\

We also assume $\alpha_h$ has a normal distribution with mean zero and variance one. We have, by the logged pdf of normal distribution, 
We have the marginal logarithm marginal likelihood for all the parameters for hospital $h$ in the safety domain satisfies 
\begin{align}\nonumber
 &-2\log \mathcal{L}(\mu_1 \dots \mu_m, \gamma_1 \dots \gamma_m, \sigma_1 \dots \sigma_m|Y_{.h}) = m\log 2\pi \\ \nonumber
 & +\log ({1}+\sum \limits_{j=1}^m \frac{\gamma_j^2}{\sigma_j^2})+\sum \limits_{j=1}^m  \log
 \sigma_j^2+\sum \limits_{j=1}^m \frac{(Y_{jh}-\mu_j)^2}{\sigma_j^2}-\frac{(\sum \limits_{j=1}^m\frac{(Y_{jh}-\mu_j)\gamma_j}{\sigma_j^2})^2}{1+\sum \limits_{j=1}^m \frac{\gamma_j^2}{\sigma_j^2}}.
\end{align} 

\subsection{Asymptotic Behavior of LVM} \label{7.1}

We start with $m=3$. Based on the marginal log-likelihood with uniform weight, we have negative marginal log-likelihood for hospital $h$ satisfies 
\begin{align}\nonumber
2\text{NMLL}_h &=  \log ({1}+\sum \limits_{j=1}^3 \frac{\gamma_j^2}{\sigma_j^2})+\sum \limits_{j=1}^3  \log \sigma_j^2+\sum \limits_{j=1}^3 \frac{(Y_{jh}-\mu_j)^2}{\sigma_j^2}-\frac{(\sum \limits_{j=1}^3\frac{(Y_{jh}-\mu_j)\gamma_j}{\sigma_j^2})^2}{1+\sum \limits_{j=1}^3 \frac{\gamma_j^2}{\sigma_j^2}}\\ \nonumber
&+3\log 2 \pi=3\log 2\pi +\log(\sigma_1^2\sigma_2^2\sigma_3^2[1+\frac{\gamma_1^2}{\sigma_1^2}+\frac{\gamma_2^2}{\sigma_2^2}+\frac{\gamma_3^2}{\sigma_3^2}]) \\ \nonumber
&+\bigg\{(Y_{1h}-\mu_1)^2(\sigma_2^2\sigma_3^2+\frac{\sigma_2^2\sigma_3^2\gamma_1^2}{\sigma_1^2}+\sigma_3^2\gamma_2^2+\sigma_2^2\gamma_3^2-\frac{\gamma_1^2\sigma_1^2\sigma_2^2\sigma_3^2}{\sigma_1^2\sigma_1^2})  \\ \nonumber
&+(Y_{2h}-\mu_2)^2(\sigma_1^2\sigma_3^2+\sigma_1^2\gamma_3^2+\sigma_3^2\gamma_1^2) +(Y_{3h}-\mu_3)^2(\sigma_1^2\sigma_2^2+\sigma_1^2\gamma_2^2+\sigma_2^2\gamma_1^2)  \\ \nonumber
&-2(Y_{1h}-\mu_1)(Y_{2h}-\mu_2)\gamma_1\gamma_2\sigma_3^2-2(Y_{2h}-\mu_2)(Y_{3h}-\mu_3)\gamma_2\gamma_3\sigma_1^2  \\ \nonumber
&-2(Y_{1h}-\mu_1)(Y_{3h}-\mu_3)\gamma_1\gamma_3\sigma_2^2 \bigg\}\bigg/(\sigma_1^2\sigma_2^2\sigma_3^2[1+\frac{\gamma_1^2}{\sigma_1^2}+\frac{\gamma_2^2}{\sigma_2^2}+\frac{\gamma_3^2}{\sigma_3^2}]). \\ \nonumber
\end{align}
Let 
\begin{align} \nonumber
A &= \sigma_1^2\sigma_2^2\sigma_3^2[1+\frac{\gamma_1^2}{\sigma_1^2}+\frac{\gamma_2^2}{\sigma_2^2}+\frac{\gamma_3^2}{\sigma_3^2}], \\ \nonumber
B_h &=(Y_{1h}-\mu_1)^2(\sigma_2^2\sigma_3^2+\sigma_3^2\gamma_2^2+\sigma_2^2\gamma_3^2) +(Y_{2h}-\mu_2)^2(\sigma_1^2\sigma_3^2+\sigma_1^2\gamma_3^2+\sigma_3^2\gamma_1^2) \\ \nonumber
& +(Y_{3h}-\mu_3)^2(\sigma_1^2\sigma_2^2+\sigma_1^2\gamma_2^2+\sigma_2^2\gamma_1^2)-2(Y_{1h}-\mu_1)(Y_{3h}-\mu_3)\gamma_1\gamma_3\sigma_2^2  \\ \nonumber
&-2(Y_{1h}-\mu_1)(Y_{2h}-\mu_2)\gamma_1\gamma_2\sigma_3^2-2(Y_{2h}-\mu_2)(Y_{3h}-\mu_3)\gamma_2\gamma_3\sigma_1^2 ,
\end{align}
then we have 
\begin{align} \label{du}
2\text{NMLL}_h = \log A + \frac{B_h}{A}+3\log 2\pi
\end{align}
and the expected negative marginal log-likelihood for $H$ hospitals (denoted by $\text{ENMLL}_H$) is
\begin{align} \label{du1}
\text{ENMLL}_H = \frac{1}{H}\sum \limits_{h=1}^H \text{NMLL}_h = \frac{1}{2}\log A + \frac{\frac{1}{H}\sum \limits_{h=1}^H  B_h}{2A}+\frac{3}{2}\log 2\pi.
\end{align}



Recall our model is
$$Y_{jh}|\alpha_h \sim N(\mu_j+\gamma_j\alpha_h, \sigma_j^2) \quad \quad 1\le j\le m,$$

by strong law of large numbers, we have
\begin{align}\label{zhang}
   \frac{1}{H}\sum \limits_{h=1}^H (Y_{ih}-\mu_i)^2 \xrightarrow{a.s.} \sigma_i^2+\gamma_i^2
\end{align}
holds for $i=1,\dots,m$. Also note that 

\begin{align}\label{deng}
\frac{1}{H}\sum \limits_{h=1}^H (Y_{ih}-\mu_i)(Y_{jh}-\mu_j) \xrightarrow{a.s.} \gamma_i\gamma_j
\end{align}
holds for $1 \le i, j \le m$.\\ 

Therefore, plug \eqref{zhang} and \eqref{deng} into $B$, 
we have 
\begin{align} \nonumber
\frac{1}{H}\sum \limits_{h=1}^H B_h &\xrightarrow{a.s.} \sum \limits _{i,j,k \in \{1,2,3\}}\big[(\sigma_i^2+\gamma_i^2)(\sigma_j^2\sigma_k^2+\sigma_j^2\gamma_k^2+\sigma_k^2\gamma_j^2)-2\gamma_i^2\gamma_j^2\sigma_k^2\big] \\ \nonumber
&=3(\sigma_1^2\sigma_2^2\sigma_3^2 + \gamma_1^2\sigma_2^2\sigma_3^2 +\sigma_1^2\gamma_2^2\sigma_3^2+\sigma_1^2\sigma_2^2\gamma_3^2)=3A
\end{align}

holds for sufficient large $H$.
Therefore, we have  $$\text{ENMLL}_H \xrightarrow{a.s.} \frac{1}{2}\log A+\frac{3}{2}(1+\log 2\pi),$$
as $H \to \infty$.

\subsection{Multivariate Normal Distribution}
\label{7.2}

Let $Z_{jh} \sim N(\mu_j, \sigma_j^2+\gamma_j^2),  1\le j\le 3, 1\le h\le H$, moreover, assume the covariance between $Z_{i.}$ and $Z_{j.}$ is $\gamma_i\gamma_j$. Then we can have the Variance co-variance matrix of $Z_{1.}, Z_{2.}, Z_{3.}$ is:\\

$$\Sigma =
\begin{bmatrix}
    \sigma_1^2+\gamma_1^2 & \gamma_1\gamma_2 & \gamma_1\gamma_3\\
    \gamma_2\gamma_1 & \sigma_2^2+\gamma_2^2 & \gamma_2\gamma_3  \\
    \gamma_3\gamma_1 & \gamma_3\gamma_2 &  \sigma_3^2+\gamma_3^2
\end{bmatrix},
$$\\

we can show that the determinant of $\Sigma$ satisfies

\begin{align}\nonumber
    |\Sigma|&=(\sigma_1^2+\gamma_1^2)[(\sigma_2^2+\gamma_2^2)(\sigma_3^2+\gamma_3^2)-\gamma_2^2\gamma_3^2]-\gamma_1\gamma_2[\gamma_1\gamma_2(\sigma_3^2+\gamma_3^2)-\gamma_1\gamma_2\gamma_3^2] \\ \nonumber
    &+\gamma_1\gamma_3[\gamma_1\gamma_2^2\gamma_3-\gamma_1\gamma_3(\sigma_2^2+\gamma_2^2)] \\ \nonumber
    &=(\sigma_1^2+\gamma_1^2)(\sigma_2^2\sigma_3^2+\sigma_3^2\gamma_2^2+\sigma_2^2\gamma_3^2)-\gamma_1^2\gamma_2^2\sigma_3^2-\gamma_1^2\gamma_3^2\sigma_2^2\\ \nonumber
    &=\sigma_1^2\sigma_2^2\sigma_3^2 + \gamma_1^2\sigma_2^2\sigma_3^2 +\sigma_1^2\gamma_2^2\sigma_3^2+\sigma_1^2\sigma_2^2\gamma_3^2=A,
\end{align}

and we have 
$$\Sigma^* =
\begin{bmatrix}
    \sigma_2^2\sigma_3^2+\sigma_3^2\gamma_2^2+\sigma_2^2\gamma_3^2 & -\gamma_1\gamma_2\sigma_3^2 & -\gamma_1\gamma_3\sigma_2^2\\
    -\gamma_2\gamma_1\sigma_3^2 & \sigma_1^2\sigma_3^2+\sigma_3^2\gamma_1^2+\sigma_1^2\gamma_3^2 & -\gamma_2\gamma_3\sigma_1^2  \\
    -\gamma_3\gamma_1\sigma_2^2 & -\gamma_3\gamma_2\sigma_1^2 &  \sigma_2^2\sigma_1^2+\sigma_1^2\gamma_2^2+\sigma_2^2\gamma_1^2
\end{bmatrix},
$$\\

let ${Z_{.h}} = [Z_{1h}, Z_{2h}, Z_{3h}]'$ and ${\mu} = [\mu_1, \mu_2, \mu_3]'$,
thus we have the negative joint log-likelihood of multivariate normal distribution for hospital $h$ satisfies

\begin{align} \nonumber
    \text{NLL}_h &\propto \log A+({Z_{.h}}-{\mu})'\Sigma^{-1}({Z_{.h}}-{\mu}) \\ \nonumber
    & =  \log A+({Z_{.h}}-{\mu})'\frac{\Sigma^*}{A}({Z_{.h}}-{\mu}).
\end{align}

Denote $C_h=({Z_{.h}}-{\mu})'\Sigma^*({Z_{.h}}-{\mu})$, by matrix calculation, we have 
\begin{align}\nonumber
    C_h&=(Z_{1h}-\mu_1)^2(\sigma_2^2\sigma_3^2+\sigma_3^2\gamma_2^2+\sigma_2^2\gamma_3^2) -2(Z_{1h}-\mu_1)(Z_{3h}-\mu_3)\gamma_1\gamma_3\sigma_2^2 \\ \nonumber
&+(Z_{2h}-\mu_2)^2(\sigma_1^2\sigma_3^2+\sigma_1^2\gamma_3^2+\sigma_3^2\gamma_1^2) +(Z_{3h}-\mu_3)^2(\sigma_1^2\sigma_2^2+\sigma_1^2\gamma_2^2+\sigma_2^2\gamma_1^2)  \\ \nonumber
&-2(Z_{1h}-\mu_1)(Z_{2h}-\mu_2)\gamma_1\gamma_2\sigma_3^2-2(Z_{2h}-\mu_2)(Z_{3h}-\mu_3)\gamma_2\gamma_3\sigma_1^2.
\end{align}

Since the only difference between $C_h$ and $B_h$ is the notation $Y,Z$.
Therefore, we have for any hospital $h$, the negative multivariate log-likelihood is equal to the negative marginal log-likelihood, since they are both negative log-likelihoods.

\subsection{generalized LVM}
\label{7.3}

We can generalize the latent variable mode with uniform weight from $3$ indicators into $m$ indicators. Note that, with a distance of $m\log 2\pi$,
\begin{align}\nonumber
2\text{NMLL}_h &=  \log ({1}+\sum \limits_{j=1}^m \frac{\gamma_j^2}{\sigma_j^2})+\sum \limits_{j=1}^m  \log \sigma_j^2+\sum \limits_{j=1}^m \frac{(Y_{jh}-\mu_j)^2}{\sigma_j^2}-\frac{(\sum \limits_{j=1}^m\frac{(Y_{jh}-\mu_j)\gamma_j}{\sigma_j^2})^2}{1+\sum \limits_{j=1}^m \frac{\gamma_j^2}{\sigma_j^2}}\\ \nonumber
&=\log \big[(1+\sum \limits _{j=1}^m \frac{\gamma_j^2}{\sigma_j^2}) \prod_{j=1}^m \sigma_j^2 \big] + \Big \{ \sum \limits _{j=1}^m \big[ (Y_{jh}-\mu_j)^2 (1+\sum \limits _{k \neq j} \frac{\gamma_k^2}{\sigma_k^2})\prod_{k \neq j} \sigma_k^2 \big] \\ \nonumber
& - 2\sum \limits_{i \neq j}\big[ (Y_{ih}-\mu_i)(Y_{jh}-\mu_j)\gamma_i \gamma_j \prod_{k \neq i, k\neq j}\sigma_k^2 \big] \Big\}/\big[(1+\sum \limits _{j=1}^m \frac{\gamma_j^2}{\sigma_j^2})\prod_{j=1}^m \sigma_j^2 \big] \\ \nonumber
\end{align}
By \eqref{zhang} and \eqref{deng}, for sufficient large $H$, we have, with $m$ indicators, 
\begin{align}\nonumber
2\text{ENMLL}_H & \xrightarrow{a.s.} \log \big[(1+\sum \limits _{j=1}^m \frac{\gamma_j^2}{\sigma_j^2})\prod_{j=1}^m \sigma_j^2 \big] + \Big \{ \sum \limits_{j=1}^m\big[(\sigma_j^2+\gamma_j^2)(1+\sum \limits _{k \neq j} \frac{\gamma_k^2}{\sigma_k^2})\prod_{k \neq j} \sigma_k^2\big]
\\ \nonumber
&-2\sum \limits_{i \neq j}\big[\gamma_i^2 \gamma_j^2 \prod_{k \neq i, k\neq j}\sigma_k^2 \big]\Big\}
/\big[(1+\sum \limits _{j=1}^m \frac{\gamma_j^2}{\sigma_j^2})\prod_{j=1}^m \sigma_j^2 \big] +{m}\log 2\pi\\ \nonumber
&=  \log \big[(1+\sum \limits _{j=1}^m \frac{\gamma_j^2}{\sigma_j^2})\prod_{j=1}^m \sigma_j^2 \big] + (S_1-S_2)/\big[(1+\sum \limits _{j=1}^m \frac{\gamma_j^2}{\sigma_j^2})\prod_{j=1}^m \sigma_j^2 \big] +m\log 2\pi.
\end{align}

Note that 


\begin{align}\nonumber
    S_1&=\sum \limits_{j=1}^m\big[(\sigma_j^2+\gamma_j^2)(1+\sum \limits _{k \neq j} \frac{\gamma_k^2}{\sigma_k^2})\prod_{k \neq j} \sigma_k^2\big] \\ \nonumber
    &=\sum \limits_{j=1}^m\big[\sigma_j^2(1+\sum \limits _{k \neq j} \frac{\gamma_k^2}{\sigma_k^2})\prod_{k \neq j} \sigma_k^2\big]+\sum \limits_{j=1}^m\big[\gamma_j^2(1+\sum \limits _{k \neq j} \frac{\gamma_k^2}{\sigma_k^2})\prod_{k \neq j} \sigma_k^2 \big]\\ \nonumber
    &=\sum \limits_{j=1}^m (\prod_{k =1}^m \sigma_k^2+\sum \limits _{k \neq j} \frac{\gamma_k^2}{\sigma_k^2}\prod_{k =1}^m \sigma_k^2+\gamma_j^2\prod_{k \neq j} \sigma_k^2)+\sum \limits_{j=1}^m (\gamma_j^2 \sum \limits _{k \neq j}\frac{\gamma_k^2}{\sigma_k^2}\prod_{k \neq j} \sigma_k^2),
\end{align}

by symmetry,
\begin{align}\nonumber
    S_1&=m(\prod_{k =1}^m \sigma_k^2+\sum \limits _{k \neq j} \frac{\gamma_k^2}{\sigma_k^2}\prod_{k =1}^m \sigma_k^2+\frac{\gamma_j^2}{\sigma_j^2}\prod_{k=1}^m \sigma_k^2)+2\sum \limits_{i \neq j}\big[\gamma_i^2 \gamma_j^2 \prod_{k \neq i, k\neq j}\sigma_k^2 \big].\\ \nonumber
    &=m(\prod_{k =1}^m \sigma_k^2+\sum \limits _{k=1}^m \frac{\gamma_k^2}{\sigma_k^2}\prod_{k =1}^m \sigma_k^2)+S_2,
\end{align}
which implies
\begin{align}\nonumber
    (S_1-S_2)/\big[(1+\sum \limits _{j=1}^m \frac{\gamma_j^2}{\sigma_j^2})\prod_{j=1}^m \sigma_j^2 \big]=m.
\end{align}

Therefore, we have, for $m$ indicators

\begin{align}\label{ppp}
\text{ENMLL}_H & \xrightarrow{a.s.}  \frac{1}{2}\log \big[ (1+\sum \limits _{j=1}^m \frac{\gamma_j^2}{\sigma_j^2})\prod_{j=1}^m \sigma_j^2 \big]+\frac{m}{2}(\log 2\pi+1).
\end{align}

Furthermore, denote the right hand side of \eqref{ppp} by $f(m)$, and the 4th moment of each $Y_i$ exists since each $Y_j$ has a normal distribution, then by Lindeberg–L\'evy central limit theorem, 

\begin{align}\nonumber
    \sqrt{H}(\frac{1}{H}\sum \limits_{h=1}^H \text{NMLL}_h-\text{ENMLL}_H)\xrightarrow{d} N(0,\boldsymbol{\sigma^2})
\end{align}

holds as $H\to \infty$, since $\text{ENMLL}_H$ convergence almost surely for every fixed positive integer $m$, and
\begin{align}\nonumber
  \boldsymbol{\sigma^2}&= \lim \limits _{H \to \infty}\frac{1}{4H}\sum \limits_{h=1}^H (2\text{NMLL}_h-2\text{ENMLL}_H)^2 \\ \nonumber
 &= \lim \limits _{H \to \infty}\frac{1}{4H}\sum \limits_{h=1}^H \Big \{2\text{NMLL}_h-m- \log \big[ (1+\sum \limits _{j=1}^m \frac{\gamma_j^2}{\sigma_j^2})\prod_{j=1}^m \sigma_j^2 \big]-m\log 2\pi\Big \}^2 \\ \nonumber
 &=  \lim \limits _{H \to \infty}\frac{1}{4H}\sum \limits_{h=1}^H\Big \{ \sum \limits _{j=1}^m \big[ (Y_{jh}-\mu_j)^2 (1+\sum \limits _{k \neq j} \frac{\gamma_k^2}{\sigma_k^2})\prod_{k \neq j} \sigma_k^2 \big] \\ \nonumber
& - 2\sum \limits_{i \neq j}\big[ (Y_{ih}-\mu_i)(Y_{jh}-\mu_j)\gamma_i \gamma_j \prod_{k \neq i, k\neq j}\sigma_k^2 \big]\Big\}^2/\big[(1+\sum \limits _{j=1}^m \frac{\gamma_j^2}{\sigma_j^2})\prod_{j=1}^m \sigma_j^2 \big]^2-m^2
\\ \nonumber
& < m^4P^{m}-m^2 <\infty,
\end{align}
where $P=\max \limits_{1 \le j \le m} E(Y_j-\mu_j)^4=\max \limits_{1 \le j \le m}3(\sigma_j^2+\gamma_j^2)^2.$
\subsection{Weighted LVM}
\label{7.4}

Let the weight matrix $W=[w_{jh}]_{m\times H}$ where $w_{jh}=w_j$, are positive constants. And
$$Y_{jh}|\alpha_h \sim N(\mu_j+\gamma_j\alpha_h, \frac{\sigma_j^2}{w_j}). \quad \quad 1\le j\le m$$
Then by definition of NMLL, we have the marginal log-likelihood for hospital $h$ satisfies
\begin{align}\nonumber
  & \quad -2\log \mathcal{L}(\mu_1 \dots \mu_m, \gamma_1 \dots \gamma_m, \frac{\sigma_1}{\sqrt{w_1}} \dots \frac{\sigma_m}{\sqrt{w_m}}|Y_{.h}) -m\log 2\pi \\ \nonumber
 &= \log ({1}+\sum \limits_{j=1}^m \frac{w_j\gamma_j^2}{\sigma_j^2})+\sum \limits_{j=1}^m  \log \frac{\sigma_j^2}{w_j}+\sum \limits_{j=1}^m \frac{w_j(Y_{jh}-\mu_j)^2}{\sigma_j^2}-\frac{(\sum \limits_{j=1}^m\frac{w_j(Y_{jh}-\mu_j)\gamma_j}{\sigma_j^2})^2}{1+\sum \limits_{j=1}^m \frac{w_j\gamma_j^2}{\sigma_j^2}} \\ \nonumber
 &= \log ({1}+\sum \limits_{j=1}^m \frac{w_j\gamma_j^2}{\sigma_j^2})+\sum \limits_{j=1}^m  w_j\log \sigma_j^2+\sum \limits_{j=1}^m \frac{w_j(Y_{jh}-\mu_j)^2}{\sigma_j^2}-\frac{(\sum \limits_{j=1}^m\frac{w_j(Y_{jh}-\mu_j)\gamma_j}{\sigma_j^2})^2}{1+\sum \limits_{j=1}^m \frac{w_j\gamma_j^2}{\sigma_j^2}} \\ \nonumber
 &\quad -\sum \limits_{j=1}^m  \log w_j -\sum \limits_{j=1}^m  (w_j-1)\log  \sigma_j^2\\ \nonumber
 & = -2\log \mathcal{L}^*(\mu_1 \dots \mu_m, \gamma_1 \dots \gamma_m, {\sigma_1} \dots \sigma_m|Y_{.h}) -\sum \limits_{j=1}^m  \log w_j -\sum \limits_{j=1}^m  (w_j-1)\log  \sigma_j^2\\ \nonumber
 &-\sum \limits_{j=1}^m w_{j} \log 2\pi.
\end{align} 
By Theorem 2,
\begin{align} \nonumber
     -\frac{2}{H}\sum \limits _{h=1}^H\log \mathcal{L}(\mu_1 \dots \mu_m, \gamma_1 \dots \gamma_m, \frac{\sigma_1}{\sqrt{w_1}} \dots \frac{\sigma_m}{\sqrt{w_m}}|Y_{.h})-m\log 2\pi\\ \nonumber \xrightarrow{a.s.}m+\log \big[(1+\sum \limits _{j=1}^m\frac{w_j\gamma_j^2}{\sigma_j^2})\prod_{j=1}^m\frac{\sigma_j^2}{w_j} \big].
\end{align}

Therefore, we have, the expected negative pseudo marginal log-likelihood ($\text{ENWMLL}_H$) for $m$ indicators satisfies

\begin{align}\nonumber
    2\text{ENWMLL}_H &\xrightarrow{a.s.} m+\log \big[(1+\sum \limits _{j=1}\frac{w_j\gamma_j^2}{\sigma_j^2})\prod_{j=1}^m\frac{\sigma_j^2}{w_j} \big]+\sum\limits_{j=1}^m  (w_j-1)\log  \sigma_j^2\\ \nonumber
    &+\sum \limits_{j=1}^m  \log w_j+\sum \limits_{j=1}^m w_{j} \log 2\pi,
\end{align}

as $H \to \infty$. Similarly, the condition for Lindeberg–L\'evy central limit theorem holds for NWMLL with every $m$.

\end{document}